Τίτλος:
# Κβαντική χημεία, μαγνητισμός και λέηζερ.

<u>Γεώργιος Λευκίδης</u>
Τμήμα Φυσικής του Πανεπιστημίου του Kaiserslautern,
Τομέας Θεωρίας Συμπυκνωμένης Ύλης.
Box 3049
67653 Kaiserslautern
Γερμανία
Τηλ.: +49 631 2053207
Φαξ: +49 631 2053907
e-mail: <u>lefkidis@physik.uni-kl.de</u>

# Κβαντική χημεία, μαγνητισμός και λέηζερ

**Περίληψη**
Με κβαντοχημικούς υπολογισμούς υψηλού επιπέδου συσχέτισης υπολογίζονται από πρώτες αρχές οι συνεισφορές των ηλεκτρικών διπόλων και τετραπόλων και των μαγνητικών διπόλων στην μη γραμμική οπτική του NiO. Το υλικό μοντελοποιείται ως ένα cluster με τεχνική διπλής εμβάπτισης και υπολογίζονται *όλες* οι μαγνητικές μεσοζωνικές *d*-καταστάσεις εξηγώντας πλήρως τα πειραματικά δεδομένα.

Γεώργιος Λευκίδης
Τμήμα Φυσικής του Πανεπιστημίου του Kaiserslautern,
Τομέας Θεωρίας Συμπυκνωμένης Ύλης.
Box 3049
67653 Kaiserslautern
Γερμανία
Τηλ.: +49 631 2053207
Φαξ: +49 631 2053907
e-mail: lefkidis@physik.uni-kl.de

# Quantum chemistry, magnetism and lasers.

**Summary**
Using highly correlational quantum chemistry we compute from first principles the contributions of the electric dipoles and quadupoles, and magnetic dipoles to the nonlinear optics of NiO. The material is modeled as a doubly embedded cluster and *all* magnetic intragap *d*-character states are calculated, thus explaining the experimental results.

George Lefkidis
Physics Department, Condensed Matter Theory Group
University of Kaiserslautern
Box 3049
67653 Kaiserslautern
Germany
Tel.: +49 631 2053207
Fax: +49 631 2053907
e-mal: lefkidis@physik.uni-kl.de

Τα τελευταία χρόνια οι ηλεκτρονικές συσκευές έχουν φτάσει σε ένα τέτοιο επίπεδο τεχνολογικής εξέλιξης, που η περαιτέρω βελτίωσή τους αποτελεί συνεχή πρόκληση. Ένα από τα βασικά συστατικά στοιχεία ενός σύγχρονου Η/Υ που όμως έχει σχεδόν αγγίξει τα θεμελιώδη όρια μεγέθους και ταχύτητας με βάση τις υπάρχουσες τεχνικές είναι τα μαγνητικά μέσα αποθήκευσης. Οι κόκκοι στην επιφάνεια των σημερινών σκληρών δίσκων είναι τόσο μικροί ώστε ο κβαντικός τους χαρακτήρας να μην μπορεί πλέον να αγνοηθεί. Πέραν τούτου οι μηχανισμοί εγγραφής και ανάγνωσης της μαγνητικής τους κατάστασης εξακολουθεί να βασίζεται στα μαγνητικά πεδία των κεφαλών γεγονός που καθιστά αδύνατους χρόνους μικρότερους από κάποια πικοδευτερόλεπτα. Για να μπορέσουμε να σχεδιάσουμε νέους ακόμα ταχύτερους σκληρούς δίσκους, θα πρέπει να καταφύγουμε στις αλληλεπιδράσεις ηλεκτρονίων-φωτονίων, οι οποίες λαμβάνουν χώρα σε χρόνους της τάξεως των φεμτοδευτερολέπτων. Με άλλα λόγια να χρησιμοποιήσουμε ακτίνες λέιζερ.

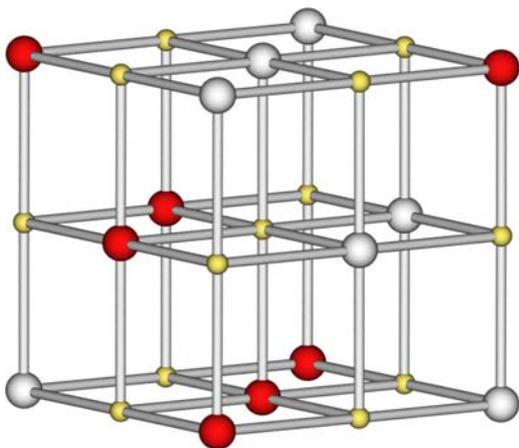

Εικ. 1: Αντισιδηρομαγνητική δομή NiO. Οι κόκκινες μπάλες δείχνουν ιόντα Ni(↑↑), οι άσπρες ιόντα Ni(↓↓) και οι μικρές κίτρινες ιόντα οξυγόνου.

Ένα πρότυπο σύστημα, ικανό να επιτελέσει έναν τέτοιο σκοπό, πρέπει να έχει τουλάχιστον δύο διαφορετικές βασικές μαγνητικές καταστάσεις και διεγερμένες καταστάσεις με σαφείς διαφορές ενέργειας, ούτως ώστε να είναι δυνατή η επιλεκτική διέγερση του συστήματος. Το τρίτο απαραίτητο χαρακτηριστικό είναι η ύπαρξη σύζευξης σπιν-τροχιάς, η οποία σε συνδυασμό με το ηλεκτρικό πεδίο του λέιζερ μπορεί να οδηγήσει σε αναστροφή μαγνήτισης[1]. Ένα τέτοιο υλικό είναι το οξείδιο του νικελίου. Κρυσταλλώνεται στο κυβικό σύστημα, και είναι αντισιδηρομαγνητικός μονωτής σε θερμοκρασία δωματίου ($T_{Néel}$ = 423° K). Τα δισθενή ιόντα του Ni βρίσκονται στην τριπλή κατάσταση (δύο μονήρη $d$-ηλεκτρόνια), τα οποία σε κάθε επίπεδο (111) του κρυστάλλου είναι διατεταγμένα με παράλληλο σπιν, ενώ τα επίπεδα μεταξύ τους παρουσιάζουν αντίθετη μαγνήτιση. Για αυτό το λόγο ο κρύσταλλος παρουσιάζει και μια μικρή παραμόρφωση (συμπίεση) κατά μήκος του άξονα [111]. Η κρυσταλλική δομή του NiO είναι πολύ καλά γνωστή, και τα πειραματικά δεδομένα περιγράφουν πλήρως τις διάφορες μαγνητικές περιοχές (εικ. 1). Το ενεργειακό του χάσμα είναι περίπου 4 eV (τα διάφορα πειραματικά δεδομένα δε συμφωνούν απόλυτα μεταξύ τους), το ενδιαφέρον είναι όμως ότι μέσα στο ενεργειακό χάσμα εμφανίζει αδιάσπαρτες διεγερμένες καταστάσεις με $d$ χαρακτήρα, οι οποίες έχουν διαπιστωθεί πειραματικά εδώ και δεκαετίες τόσο για το εσωτερικό του (bulk)[2], όσο και για την επιφάνεια (001)[3]. Οι διεγέρσεις σε αυτές τις καταστάσεις είναι απαγορευμένες σε προσέγγιση ηλεκτρικών διπόλων γι'αυτό και προκαλούν πολύ ασθενείς φασματοσκοπικές απορροφήσεις.

## 1. Κβαντική Χημεία

Το σημαντικότερο εργαλείο για την περιγραφή συστημάτων από πρώτες αρχές αποτελεί η εξίσωση του Schrödinger. Ο συνήθης τρόπος λύσης της είναι μέσω της προσέγγισης Hartree-Fock. Σε αυτήν επιλύουμε το κβαντοχημικό πρόβλημα λαμβάνοντας υπόψη μας μόνο ένα μέρος από τις αλληλεπιδράσεις μεταξύ των ηλεκτρονίων, και καταλήγουμε σε ένα σύνολο μονοηλεκτρονικών μοριακών τροχιακών από τα οποία μερικά είναι κατειλημμένα από ηλεκτρόνια ενώ μερικά παραμένουν κενά. Τα κατειλημμένα τροχιακά τα συγκεντρώνουμε σε μια ορίζουσα Slater, η οποία μας δίνει την ολική πολυηλεκτρονική κυματοσυνάρτηση του συστήματος ικανοποιώντας ταυτόχρονα τη φερμιονική απαίτηση αντισυμμετρικότητας της. Ακολούθως, και ανάλογα με το χημικό πρόβλημα, υπολογίζουμε τις συσχετίσεις, πολύ συχνά με κάποια μέθοδο αλληλεπίδρασης διαμόρφωσης. Έτσι ξεκινώντας από την ορίζουσα Slater που μας δίνει η μέθοδος Hartree-Fock φτιάχνουμε περαιτέρω ορίζουσες αντικαθιστώντας ένα ή περισσότερα κατηλειμμένα μοριακά τροχιακά με κενά. Με αυτές σχηματίζουμε έναν γραμμικό συνδυασμό και ελαχιστοποιούμε την ολική ενέργεια με τη μέθοδο των μεταβολών. Κατ'αυτόν τον τρόπο μπορούμε να πετύχουμε καλύτερη περιγραφή τόσο της βασικής όσο και των διεγερμένων καταστάσεων. Ανάλογα με τον αριθμό των διεγερμένων ηλεκτρονίων που επιτρέπουμε (δηλ. πόσα κατειλημμένα τροχιακά αντικαθιστούμε με κενά), έχουμε διάφορες μεθόδους, όπως CIS (configuration interaction with single excitations) αν έχουμε μόνο απλές διεγέρσεις, CISD (singles and doubles) αν επιτρέψουμε έως και διπλές διεγέρσεις, CISDT (singles, doubles and triples) μέχρι τριπλές διεγέρσεις, κ.ο.κ. Αν επιτρέψουμε όλες τις δυνατές διεγέρσεις τότε, έχουμε την πλήρη αλληλεπίδραση διαμόρφωσης (full CI), που δίνει τα καλύτερα δυνατά αποτελέσματα για κάθε δεδομένο σύνολο βάσης. Δυστυχώς όμως ο αριθμός των οριζουσών αυξάνει εκθετικά με το σύνολο των διεγέρσεων, καθιστώντας παρόμοιους υπολογισμούς μη πραγματοποιήσιμους παρεκτός για πολύ μικρά μόρια.

Το οξείδιο του νικελίου είναι ένα υλικό για το οποίο είναι γνωστή η ανεπάρκεια της προσέγγισης Hartree-Fock, αφού οι συσχετίσεις του συστήματος ακόμα και σε εντοπισμένο επίπεδο είναι αρκετά εκτεταμμένες. Εκτός αυτού ο πολυοριζουσιακός χαρακτήρας ακόμα και των απλώς διεγερμένων καταστάσεων είναι σημαντικός. Με άλλα λόγια, η χρήση μιας μόνο ορίζουσας για την περιγραφή των διεγερμένων καταστάσεων δεν είναι αρκετή. Στην απλούστερη περίπτωση πρέπει κανείς να λάβει υπόψη του τουλάχιστον δύο ορίζουσες Slater, οι οποίες συμμετέχουν στην βασική κατάσταση με ίσο σχεδόν ποσοστό. Μια απλουστευμένη εικόνα μπορεί να έχει κανείς αν σκεφτεί την ηλεκτρονική διαμόρφωση των οκτώ *d* ηλεκτρονίων έκαστου νικελίου στο οκταεδρικό πεδίο υποκαταστατών των οξυγόνων, από τα οποία έξι καταλαμβάνουν ανά δύο τις τρεις χαμηλότερες $t_{2g}$ καταστάσεις και τα υπόλοιπα δύο με παράλληλο σπιν τις υψηλότερες $e_g$. Οπότε για την μονοηλεκτρονική διέγερση $t_{2g} \leftrightarrow e_g$ υπάρχουν συνολικά έξι ισοπίθανες δυνατότητες, λογικό λοιπόν μια μοναδική ορίζουσα να περιγράφει ελλειπώς το σύστημα.

## 2. Μη γραμμική οπτική

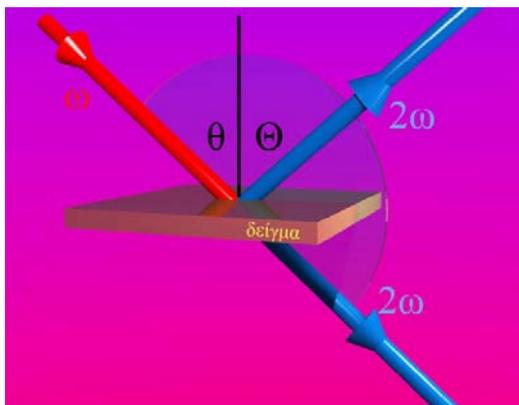

Εικ. 2: Προσπίπτουσα, ανακλώμενη και διαθλώμενη δέσμη φωτός, με συχνότητες ω και 2ω.

Ο καλύτερος τρόπος χαρακτηρισμού υλικών είναι αναμφισβήτητα, εφόσον αυτό είναι δυνατόν, η χρήση οπτικών μέσων, λόγω της αμεσότητάς τους, της ταχύτητάς τους και του ότι δεν καταστρέφουν το δείγμα (εικ. 2). Η χρησιμοποίηση μη γραμμικής οπτικής μας επιτρέπει δε να έχουμε καλύτερη ανάλυση από ένα φάσμα απορρόφησης ή διχρωισμού μόνο, αφού πρόκειται για φαινόμενο ανώτερης τάξης και κατά συνέπεια με μεγαλύτερη διακριτική επιλεκτικότητα. Αν φωτίσουμε ένα υλικό με λέιζερ συχνότητας ω, τότε για να υπολογίσουμε την ένταση του ανακλούμενου ή διαθλούμενου φωτός θα πρέπει να λύσουμε το σύστημα των συζευγμένων εξισώσεων Maxwell για την απλή (ω) και τη διπλασιασμένη συχνότητα (2ω), όπου θα έχουμε προσθέσει τους πηγαίους όρους που περιγράφουν την απόκριση του υλικού:

$$\mathbf{S} = -\omega_s^2 \mu_0 \mathbf{P}_s - i\omega_s \mu_0 \nabla \times \mathbf{M}_s + \omega_s^2 \mu_0 \nabla \cdot \ddot{\mathbf{Q}}_s$$

όπου η πολωσιμότητα **P**, η μαγνήτιση **M** και η πολωσιμότητα τετραπόλων $\vec{\vec{Q}}$ μπορούν να αναπτυχθούν σε δυναμοσειρές Taylor ως προς το ηλεκτρικό (ή και μαγνητικό) πεδίο της προσπίπτουσας φωτινής δέσμης.

$$\mathbf{P_s} = \varepsilon_0 \vec{\vec{\chi}}^{ee} \mathbf{E} + \varepsilon_0 \vec{\vec{\chi}}^{eee} : \mathbf{E}^2 + \varepsilon_0 \vec{\vec{\chi}}^{eeee} : \mathbf{E}^3 + \ldots$$

όπου **E** η ένταση του ηλεκτρικού πεδίου του προσπίπτοντως φωτός, και $\vec{\vec{\chi}}$ οι τανυστές ηλεκτρικής επιδεκτικότητας νιοστής τάξης. Ο δεύτερος όρος, ο οποίος ταλαντούται με διπλάσια συχνότητα, είναι υπεύθυνος για την απόκριση δεύτερης αρμονικής και μπορεί να επεκταθεί ώστε να εξαρτάται όχι μόνο από το ηλεκτρικό αλλά και από το μαγνητικό πεδίο, καθώς επίσης να περιλαμβάνει μεταβάσεις τόσο ηλεκτρικών διπόλων, όσο και ανώτερης τάξης[5]. Υπολογίζουμε τα στοιχεία μήτρας ηλεκτρονικών μεταβάσεων ξεκινώντας από την εξίσωση του Schrödinger, όπου η ορμή του ηλεκτρονίου εμφανίζεται επαυξημένη κατά το διανυσματικό δυναμικό Α του λέηζερ (minimal Hamiltonian) στη βαθμίδα Coulomb επί το φορτίο του ηλεκτρονίου:

$$\hat{H} = \frac{(\hat{p} - q\hat{A})^2}{2m} = \frac{\hat{p}^2}{2m} - \frac{q\hat{p} \cdot \hat{A}}{m} + \underbrace{\frac{q^2 \hat{A}^2}{2m}}_{\approx 0}$$

όπου $\hat{H}$ η χαμιλτωνιανή, $\hat{p}$ ο τελεστής ορμής, q το φορτίο και m η μάζα του ηλεκτρονίου. Κατόπιν αναπτύσσουμε σε δυναμοσειρά Taylor το δεύτερο όρο, τον οποίο θεωρούμε ως μια μικρή διαταραχή, και υπολογίζουμε κατ'αυτόν τον τρόπο τα στοιχεία μετάβασης. Στην πιο συνηθισμένη προσέγγιση των ηλεκτρικών διπόλων, κρατάμε μόνο τον πρώτο όρο της δυναμοσειράς. Αν λάβουμε υπόψη μας και τον επόμενο όρο, τότε με λίγα μαθηματικά, καταλήγουμε σε ακόμα δύο προσθετέους, και τελικά τα στοιχεία διαταραχής γίνονται:

$$\mathbf{V}^e_{\alpha\beta} = -q\mathbf{d}_{\alpha\beta}$$

$$\mathbf{V}^m_{\alpha\beta} = -\frac{q}{2m}\mathbf{L}^{n \times a}_{\alpha\beta}$$

$$\vec{\vec{V}}^q_{\alpha\beta} = -q\vec{\vec{Q}}^{n,a}_{\alpha\beta}$$

όπου **d**, **L** και $\vec{\vec{Q}}$ είναι τα στοιχεία μετάβασης μεταξύ των καταστάσεων *α* και *β*, **n** η διεύθυνση διάδοσης του φωτός, και **a** η διεύθυνση πόλωσης. Ο εκθέτης του **V** δείχνει τη φύση της μετάβασης: ηλεκτρικά δίπολα (e), μαγνητικά δίπολα (m) ή ηλεκτρικά τετράπολα (q). Τα δύο τελευταία συνεισφέρουν κατ'ουσίαν στον ίδιο τανυστή δεύτερης τάξης, τα ηλεκτρικά τετράπολα (q) αντιστοιχούν στο συμμετρικό του σκέλος, ενώ τα μαγνητικά δίπολα (m) στο αντισυμμετρικό. Παρόλο που οι μεταβάσεις αυτές είναι αρκετές τάξεις μεγέθους ασθενέστερες από αυτές εξαιτίας των ηλεκτρικών διπόλων, υπάρχει τουλάχιστον μια περίπτωση όπου είναι εξαιρετικά σημαντικές: όταν οι δεύτερες είναι απαγορευμένες λόγω συμμετρίας. Μια τέτοια περίπτωση είναι οι μεταβάσεις που αφορούν διέγερση ηλεκτρονίων από ένα *d* τροχιακό σε ένα άλλο, γιατί τα ηλεκτρικά δίπολα απαιτούν αλλαγή ισοτιμίας (parity) της κυματοσυνάρτησης, δηλ. gerade σε ungerade και τούμπαλιν (κανόνας επιλογής του Laporte). Έτσι για παράδειγμα, σε ένα οκταεδρικό κρυσταλλικό πεδίο με διαχωρισμό των εκφυλισμένων πέντε *d* τροχιακών σε τρία $t_g$ και δύο $e_g$ τα οποία έχουν μεν διαφορετική ενέργεια, όμως μόνο μετάβαση ανώτερης τάξης από το ένα στο άλλο είναι επιτρεπτή.

Για να υπολογίσουμε την ηλεκτρική επιδεκτικότητα δεύτερης τάξης ($\vec{\vec{\chi}}^{(2\omega)}$) ξεκινούμε από την εξίσωση του Liouville και θεωρούμε την επίδραση του ηλεκτρομαγνητικού πεδίου του λέηζερ ως μια μικρή διαταραχή του συστήματος, του οποίου έχουμε προηγουμένως υπολογίσει τις στατικές κυματοσυναρτήσεις τόσο της βασικής όσο και των διεγερμένων καταστάσεων που μας ενδιαφέρουν ελλείψει εξωτερικών διαταραχών. Κατόπιν υπολογίζουμε τα στοιχεία της μήτρας μετάβασης μεταξύ τους, τα οποία χρησιμοποιούμε στον τύπο [4,5].

$$\chi_{ijk}^{lmn} \propto \frac{\rho_0}{\varepsilon_0} \sum_{\alpha,\beta,\gamma} \langle \alpha | d_i^l | \beta \rangle \overline{\langle \beta | d_j^m | \gamma \rangle \langle \gamma | d_k^n | \alpha \rangle} \times \frac{\dfrac{f(E_\gamma)-f(E_\beta)}{E_\gamma - E_\beta - \hbar\omega + i\omega\Gamma} - \dfrac{f(E_\gamma)-f(E_\beta)}{E_\beta - E_\alpha - \hbar\omega + i\omega\Gamma}}{E_\gamma - E_\alpha - 2\hbar\omega + 2i\omega\Gamma}$$

όπου χ είναι η ηλεκτρική επιδεκτικότητα δεύτερης τάξης, οι δείκτες *l*, *m* και *n* είναι η φύση του στοιχείου μήτρας (ηλεκτρικό δίπολο, μαγνητικό δίπολο ή ηλεκτρικό τετράπολο, e, m και q αντίστοιχα), *i*, *j* και *k* οι καρτεσιανές συντεταγμένες των αντίστοιχων μεταβάσεων, $\rho_0$ η ηλεκτρονική πυκνότητα, $\langle \alpha | d_i^m | \beta \rangle$ τα στοιχεία μήτρας μετάβασης, $E_\alpha$ η ενέργεια της κατάστασης *α*, *f(E)* η συνάρτηση ηλεκτρονικής κατανομής και Γ ο συντελεστής απόσβεσης (η μόνη παράμετρος που δεν μπορεί να υπολογιστεί με το χρησιμοποιούμενο κβαντοχημικό μοντέλο, και στον οποίο δόθηκαν τιμές που εμπειρικά αναπαράγουν την πειραματική διαπλάτυνση των γραμμών απορρόφησης, Γ ~ 0.1 eV). Η μπάρα πάνω από τα τα στοιχεία μετάβασης σημαίνουν συμμετρικοποίηση. Ο λόγος είναι η υφιστάμενη συμμετρία Kleinmann, ότι δηλαδή δεν είμαστε σε θέση να διαχωρίσουμε ποιο από τα δύο ταυτόχρονως προσπίπτοντα φωτόνια ίσης ενέργειας προκαλεί ποια από τις μεταβάσεις. (Η άνωθι συνάρτηση δεν περιλαμβάνει διορθώσεις τοπικών πεδίων λόγω υλικού – local field corrections – και ως τούτου περιγράφει μόνο την σχετική ένταση του μετρούμενου σήματος.) Αφού υπολογίσουμε όλα τα στοιχεία του $\vec{\chi}^{(2\omega)}$, είμαστε σε θέση να υπολογίσουμε και την ένταση της δεύτερης αρμονικής

$$I^{(2\omega)} \propto \left[ f(\varphi,\vartheta) \vec{\chi}^{(2\omega)} F(\Phi,\Theta) \right]^2.$$

Οι συναρτήσεις *f(φ,θ)* και *F(Φ,Θ)* εκατέρωθεν του $\chi^{(2\omega)}$ περιγράφουν την πειραματική γεωμετρία της προσπίπτουσας και ανακλώμενης (διαθλώμενης) δέσμης αντιστοίχως (εμπεριέχουν ημίτονα και συνημίτονα κατεύθυνσης και συντελεστές Fresnel) και ο υπολογισμός τους είναι θέμα απλής τριγωνομετρίας.

## 3. Υπολογιστικό μοντέλο

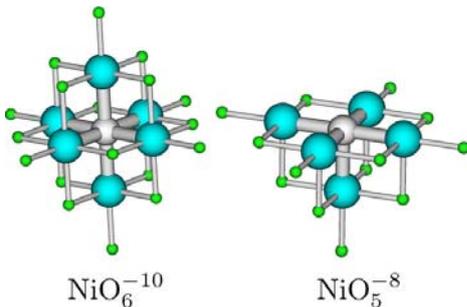

Εικ. 3: Δομές που χρησιμοποιήθηκαν για την αναπαράσταση του εσωτερικού του κρυστάλλου (αριστερά) και της επιφάνειας [001] (δεξιά). Οι μικρές πράσινες σφαίρες είναι η στοιβάδα των αποτελεσματικών δυναμικών.

Για την περιγραφή του υλικού χρησιμοποιήσαμε δύο cluster, $(NiO_6)^{-10}$ και $(NiO_5)^{-8}$ για το εσωτερικό του (bulk) και την επιφάνεια (001) αντίστοιχα (εικ. 3). Τα clusters εμβαπτίστηκαν πρώτα σε μια στοιβάδα αποτελούμενη από αποτελεσματικά δυναμικά (effective core potentials) τα οποία περιγράφουν τα ιόντα νικελίου που περιβάλλουν άμεσα το κάθε cluster, και κατόπιν σε ένα πλέγμα σημειακών φορτίων (15×15×15 και 15×15×8 για το εσωτερικό και την επιφάνεια αντίστοιχα) που περιγράφουν το πεδίο Madelung. Το σύνολο βάσης ήταν κάθε φορά το Los Alamos για το νικέλιο και το SBKJC για τα οξυγόνα[6], και επιλέχτηκαν γιατί δίνουν τα καλύτερα αποτελέσματα όσον αφορά τις ενεργειακές στάθμες αλλά και τη συμμετρία των πολυηλεκτρονικών καταστάσεων (απαραίτητο στοιχείο για την ανάλυση και έλεγχο της ορθότητας της συμμετρίας της επιδεκτικότητας $\chi^{(2\omega)}$). Οι κβαντοχημικοί υπολογισμοί έγιναν με τη μέθοδο multiconfigurational complete active space (MC-CAS) και συσχετίστηκαν όλα τα μοριακά τροχιακά με *d* χαρακτήρα του Ni. Η μέθοδος αυτή είναι παρόμοια με την αλληλεπίδραση διαμόρφωσης, όπου όμως κατά την διαδικασία ελαχιστοποίησης της ενέργειας δεν μεταβάλουμε μόνο τους συντελεστές με τους οποίους εμφανίζονται στον γραμμικό συνδυασμό οι ορίζουσες Slater αλλά επιτρέπουμε και «χαλάρωση» των αρχικών μοριακών τροχιακών. Κατά αυτόν τον τρόπο ήμασταν σε θέση να υπολογίσουμε και τις 25 μεσοζωνικές καταστάσεις του υλικού μας, τόσο τις τριπλές (triplets που είναι οι μαγνητικές) όσο και τις απλές (singlets). Σε αυτό το σημείο αξίζει να σημειώσουμε ότι οι μέθοδοι συναρτησιακής πυκνότητας (density functional theory, DFT) είναι μεν σε θέση να υπολογίσουν καλύτερα την περιοδικότητα του συστήματος όσον αφορά το ενεργειακό χάσμα, αλλά μέχρι σήμερα αποτυγχάνουν πλήρως στην περιγραφή των μεσοζωνικών καταστάσεων, εν αντιθέσει με τις μεθόδους πραγματικού χώρου (και όχι αντίστροφου) οι οποίες προβλέπουν με ακρίβεια τα πειραματικά δεδομένα[7].

## 4. Αποτελέσματα

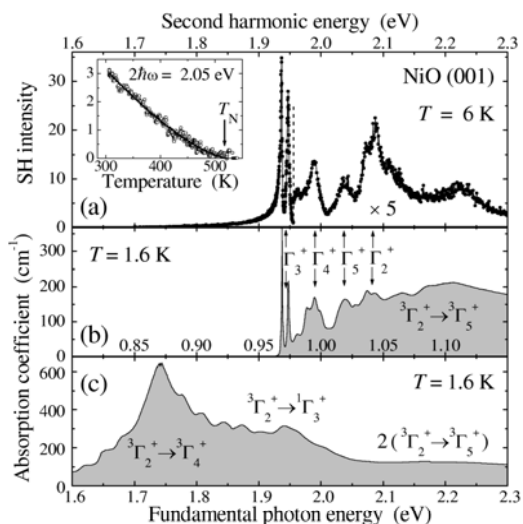

Εικ. 4: Φάσμα δεύτερης αρμονικής και γραμμικής απορρόφησης για κάθετη πρόσπτωση και ανάκλαση στο NiO. Το ένθετο δείχνει την θερμοκρασιακή εξάρτηση και καμπύλη παλινδρόμησης $I(2\omega) \propto (1 - T/T_N)^{2n\beta}$. Με την ευγενική άδεια των Fiebig *et al.*[8]

Ένα από τα βασικά χαρακτηριστικά της δεύτερης αρμονικής είναι ότι είναι απαγορευμένη σε κεντροσυμμετρικά υλικά. Ως εκ τούτου, το NiO λόγω της οκταεδρικής του συμμετρίας δεν θα έπρεπε να δίνει καθόλου σήμα, εν αντιθέσει με τα πειραματικά δεδομένα (εικ. 4)[8]. Για την εξήγηση του φαινομένου αυτού, έχουν δοθεί μέχρι στιγμής διάφορες ερμηνείες: α) ότι το σήμα προέρχεται πρωτίστως από την επιφάνεια όπου αίρεται το κέντρο συμμετρίας[9], β) ότι η σύζευξη σπιν-τροχιάς ελαττώνει τη συμμετρία (πρέπει να καταφύγει κανείς σε διπλές ομάδες σημείου για την περιγραφή του συστήματος)[10], γ) η συνεισφορά μεταβάσεων ανώτερης τάξης[5], και δ) η τοπική παραμόρφωση του κρυστάλλου, τόσο λόγω φωνονίων, όσο και λόγω της μόνιμης κρυσταλλικής παραμόρφωσης κατά μήκος του άξονα [111][11]. Όλοι οι παραπάνω λόγοι μπορούν να συνεισφέρουν, εντούτοις μόνο η ηλεκτρονικές μεταβάσεις λόγω μαγνητικών διπόλων μπορούν να εξηγήσουν τη θέση και την ένταση των παρατηρούμενων κορυφών. Οι υπόλοιποι από μόνοι τους, με βάση τουλάχιστον τα σημερινά υπολογιστικά δεδομένα, δεν αρκούν, αν και σε συνδυασμό μπορούν να διαφοροποιήσουν και να ενισχύσουν το παρατηρούμενο σήμα. Η σύζευξη σπιν-τροχιάς μειώνει τη συμμετρία, αλλά η μίξη τροχιακών με διαφορετικά σπιν, είναι της τάξεως μόλις του 3%. Πέραν τούτου η κυριότερη πειραματική κορυφή (γύρω στο 0.95 eV) αντιστοιχεί σε μετάβαση από τη βασική στην πρώτη μεσοζωνική *d*-κατάσταση του εσωτερικού (bulk). Όλες οι μεσοζωνικές καταστάσεις έχουν *d* χαρακτήρα, οπότε ακόμα και η οιαδήποτε μίξη τους, δεν θα μπορούσε να οδηγήσει σε επιτρεπτή μετάβαση μέσα στην προσέγγιση ηλεκτρικών διπόλων, αφού παραβιάζεται ο κανόνας επιλογής του Laporte. Οι τοπικές παραμορφώσεις δεν παρουσιάζουν αυτό το πρόβλημα. Ακόμα και στο απλούστερο γραμμικό μοντέλο φωνονίων, η τοπική συμμετρία μειώνεται, ανάλογα με την ιδιοκατάσταση δόνησης. Μοναδική εξαίρεση αποτελούν τα ακουστικά φωνόνια στο κέντρο της ζώνης Brillouin, όπου αντιστοιχούν απλώς σε ταυτόχρονη μετακίνηση όλων των ατόμων προς την ίδια κατεύθυνση. Τα οπτικά φωνόνια αντιθέτως, ακόμα και σε μεγάλα μήκη κύματος, προκαλούν σημαντική τοπική παραμόρφωση καθότι τα ανιόντα κινούνται προς αντίθετη κατεύθυνση από τα κατιόντα (στην περίπτωση κρυστάλλου με δομή NaCl όπως στο NiO).

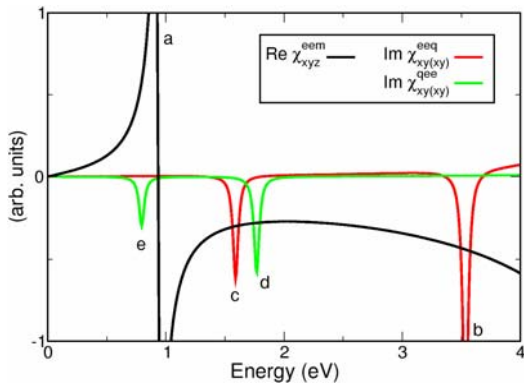

Εικ. 5: Υπολογισμένη επιδεκτικότητα δεύτερης τάξης στο εσωτερικό του NiO. Η μαύρη γραμμή είναι απόκριση λόγω μαγνητικών διπόλων, η κόκκινη και πράσινη λόγω ηλεκτρικών τετραπόλων.

Οι Fiebig et al. στα αποτελέσματα του 2001 (εικ. 4) βρίσκουν περισσότερες κορυφές δεύτερης αρμονικής. Οι ίδιοι ήδη από τότε απέδωσαν το σήμα περί τα 1 eV με βάση απλούς συλλογισμούς συμμετρίας σε μεταβάσεις μαγητικών διπόλων. Έχοντας πλέον στα χέρια μας έναν πλήρη ab initio υπολογισμό για το υλικό είμαστε σε θέση να ελέγξουμε αυτήν την υπόθεση. Πράγματι, το ισχυρότερο σήμα, κατ'απόλυτη σχεδόν συμφωνία με τα πειραματικά δεδομένα, εμφανίζεται στην περιοχή γύρω στα 1 eV και οφείλεται όντως σε μαγνητικά δίπολα (εικ. 5). Με παρόμοιο τρόπο μπορούμε να υπολογίσουμε και το σήμα της επιφάνειας (το οποίο όμως εμφανίζει την πρώτη κορυφή στα 0.5 eV και επομένως μπορεί με ασφάλεια να αποκλειστεί ως βασική πηγή του σήματος).

## 5. Σύνοψη

Στο παρόν άρθρο επιχειρίσαμε με μια σύντομη παρουσίαση αποτελεσμάτων κβαντικής χημείας υψηλού επιπέδου συσχετίσεων με πρακτική εφαρμογή στην σύγχρονη έρευνα. Δείξαμε ότι η κβαντική χημεία μπορεί να αποτελέσει ένα βασικό εργαλείο ακόμα και σε εκτεταμένα συστήματα, όπως το NiO, και να εξηγήσει τελεσίδικα πειραματικά δεδομένα που εν πρώτοις εκπλήττουν. Συγκεκριμένα δείξαμε ότι είναι δυνατόν να υπολογίσει κανείς με ακρίβεια τις μεσοζωνικές καταστάσεις ενός μονωτικού αντισιδηρομαγνητικού υλικού με βάση τις οποίες ερμηνεύεται το φάσμα μη γραμμικής οπτικής του υλικού, αποδεικνύοντας κατ'αυτόν τον τρόπο ότι κατεξοχήν «χημικές» θεωρητικές μέθοδοι συνδέονται άμεσα και με κατεξοχήν «πεδία» κλασικής φυσικής.